\documentclass{elsart}

\usepackage{natbib,amsmath,graphicx,epsfig,lineno}
\usepackage{color}
\newcommand{\dd}{\textrm{d}}

 \definecolor{darkblue}{rgb}{0.0, 0.0, 0.55}

\usepackage{setspace}
\usepackage{amssymb,amsfonts,amsmath}
\journal{Journal}

\begin{document}
\begin{frontmatter}



\title{Random patterns in fish schooling enhance alertness: a hydrodynamic perspective}

\author[1]{Usama Kadri\corauthref{cor}}, \author[2]{ Franz Br\"{u}mmer}, \author[3]{Anan Kadri} 
\corauth[cor]{ukadri@mit.edu} 

\address[1]{Department of Mathematics, Massachusetts Institute of Technology, Cambridge, Massachusetts 02139}
\address[2]{Institute of Biology, University of Stuttgart, Pfaffenwaldring 57, 70569 Stuttgart, Germany}
\address[3]{Faculty of Agricultural Sciences, Institute of Plant Breeding, Seed Science and Population Genetics, University of Hohenheim, Stuttgart, Germany}

\begin{abstract}
One of the most highly debated questions in the field of animal swarming and social behaviour, is the collective random patterns and chaotic behaviour formed by some animal species, in particular if there is a danger. Is such a behaviour beneficial or unfavourable for survival? Here we report on one of the most remarkable forms of animal swarming and social behaviour - fish schooling - from a hydrodynamic point of view. We found that some fish species do not have preferred orientation and they swarm in a random pattern mode, despite the excess of energy consumed. Our analyses, which includes calculations of the hydrodynamic forces between slender bodies, show that such a behaviour enhances the transfer of hydrodynamic information, and thus enhances the survivability of the school. These findings support the general hypothesis that a disordered and non-trivial collective behaviour of individuals within a nonlinear dynamical system is essential for optimising transfer of information - an optimisation that might be crucial for survival.

\end{abstract}
\end{frontmatter}

\section{Introduction}
The concurrent movement of fish in a school involves significant hydrodynamic interactions. The relative longitudinal and lateral distances and velocities between the fish, as well as their relative lengths and cross-sectional areas determine the magnitude of the hydrodynamic forces and moments involved \citep{Kadri:2005,  Rattanasiri:2014, Kadri:2014}, which in turn affect the school overall manoeuvrability \citep{Partridge:1979}. It is not the aim of the current study to discuss how information due to a sudden movement is (physiologically) transferred among the school members in terms of sensory systems \citep{Partridge:1980}, environmental effects \citep{Killen:2007}, or aerobic capacity \citep{Killen:2011}. In this respect, the overall manoeuvrability of a given school is dependant on the instantaneous school pattern (structure) mode which dictates to leading order, the hydrodynamic interactions. 
\begin{figure}
\centering{ \epsfig{figure=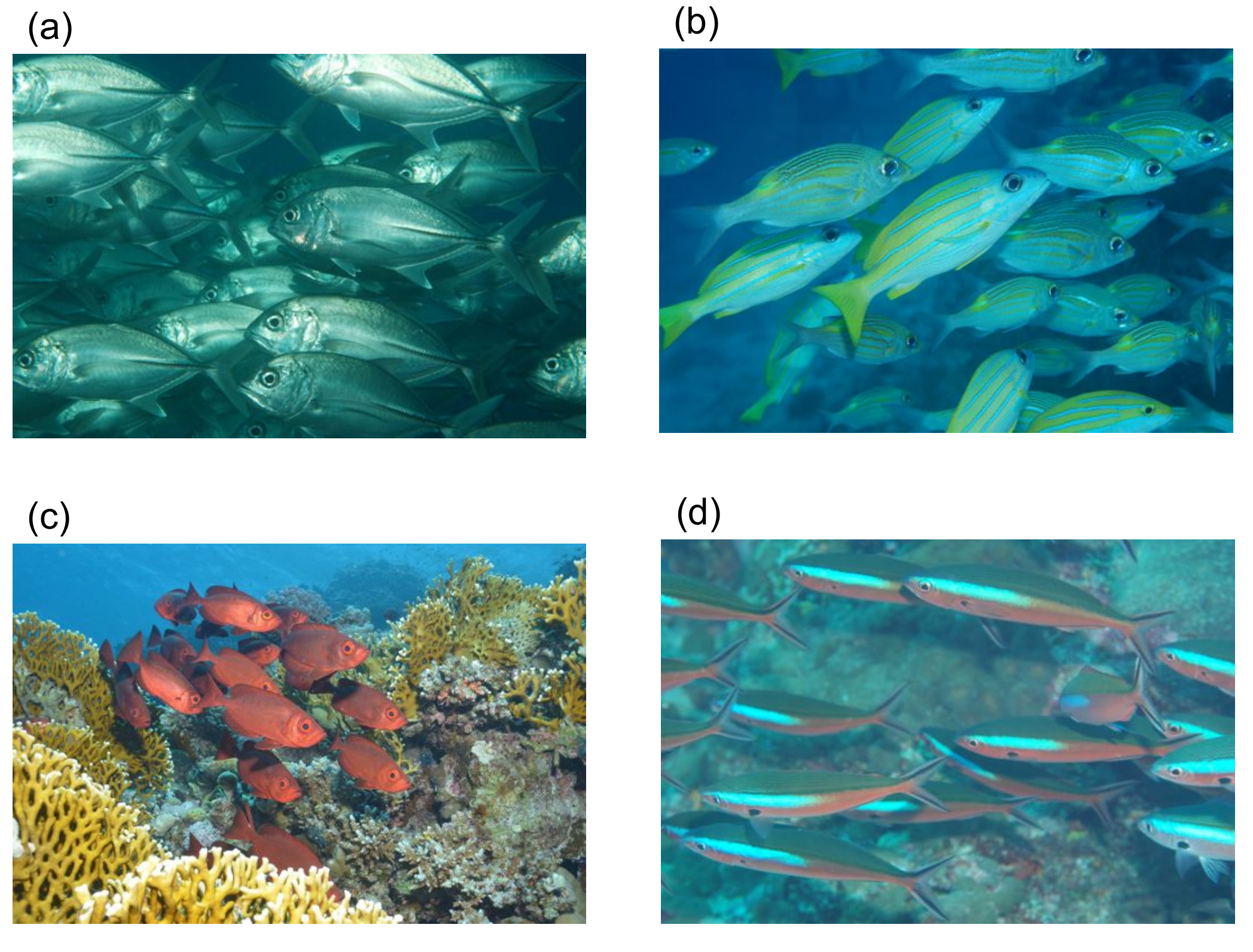,width=\linewidth}}
\caption{Examples for schools of fishes swimming in random pattern mode: (a) Jack Caranx sp. (60 cm) ; (b) Bluelined snapper Lutjanus kasmira \& Yellowspot emperor Gnathodentex aurolineatus (35 cm / 24 cm); (c) Goggle-eye Priacanthus hamrur (40 cm); and (d) Bluestreak fusilier Pterocaesio tile (25 cm). (Photos by F. Br\"{u}mmer)}
\end{figure}
\begin{figure}
\centering{ \epsfig{figure=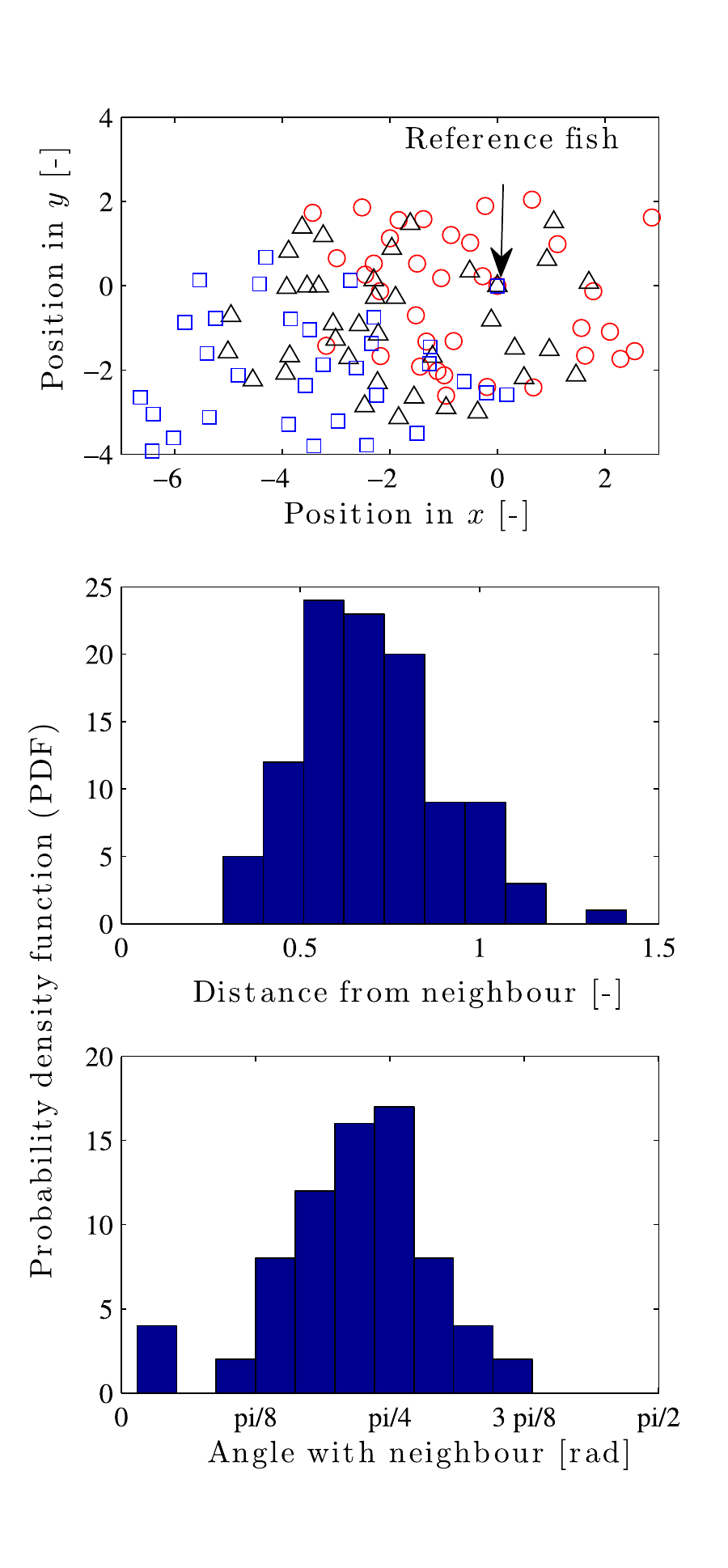,width=10cm}}
\caption{{\bf Top}: distribution of a school of Jack Caranx sp. (60 cm); at time 00.04.44 (circles), 00.05.44 (star), 00.06.59 (square). {\bf Middle}: probability density function (PDF) of the distance between each fish and the closest upper downstream neighbour. {\bf Bottom}: PDF of the angle between each fish and the closest upper downstream neighbour. All dimensionless quantities were normalized with respect to the mean fish length. }
\end{figure}

Although it has been suggested that fish might be found to swim in a diamond-shape pattern to increase hydrodynamic efficiency \citep{Weihs:1983}, or other preferred orientations and angles, observations (Fig. 1) and analyses (Fig. 2) of aerial photographs and videos of different schools of fishes (Jacks, blue-lined snapper, yellow-spot emperor, goggle-eye, and bluestreak fusilier) reveal random-shape patterns instead. The supporting theoretical analysis we present here show that swimming in random pattern modes increases the mean hydrodynamic forces by a factor of two to five, depending on school size, which in turn decreases the response time of fish due to a sudden change of movement in neighbouring fish and enhances the overall manoeuvring of the school. An increased energy consumption that enhances the manoeuvring efficiency is thus essential for survival especially amongst smaller fish that cannot escape fast enough from predators. 
\section{Methods}
\subsection{Hydrodynamic calculations}
The model by \cite{Weihs:1983} accurately predicts diamond-shape pattern modes especially for relatively large fish or dolphins \citep{Weihs:2004, Kadri:2005}, and for different types of fish preferred orientations might be identified. However, for smaller fish (e.g. Jack Caranx sp., 60 cm) the school pattern-shapes were found to be random; especially when fish encounter danger (e.g. due to the presence and sudden movement of scuba-divers) their behaviour becomes more disordered within the school; at any given instant the relative distances and angles between neighbouring fish fail to form ordered patterns, as we observed (Fig. 2). It is observed that the probability density function (PDF) of the relative distances and angles are \textit{Gaussian}, indicating continuous random variables. This observation raises the question whether random school patterns and disordered behaviour, which are probably due to a natural `panic' reflex \citep{Hamilton:1971}, are beneficial or unfavourable for survival. 

In order to evaluate the effect of random school patterns, we carried out a theoretical analysis, based on the studies by \cite{Tuck:1974}, and \cite{Wang:1975} who investigated the hydrodynamic interactions between two submerged slender bodies of revolution at various separation distances. For the sake of brevity, the actual motion of each fish in the school is now translated into the motion of a slender ellipsoid  with $d/{L}=\epsilon$, where $d$ and $L$ are the maximum lateral and longitudinal dimensions of the body, and $\epsilon$ is assumed to be small. On this basis, an approximate solution is sought for the hydrodynamic quantities of interest. Each two streamlined bodies move through an ideal fluid with constant velocities $U_i$ and $U_j$ along parallel paths. The relative positions of the two bodies change in time as a quasi--steady approximation, where each position is calculated individually. The two bodies are separated by a lateral distance, $\eta_{ij}$, and fore--and--aft distance, $\xi_{ij}$, which is a function of time $t$. For each two bodies we define two coordinate systems, $(x_i,y_i,z_i)$ fixed on body $i$ and $(x_j,y_j,z_j)$ fixed on the upper upstream neighbour, body $j$, which are related to the fixed coordinate system $(x_0,y_0,z_0)$ so that 
\begin{equation} \label{eq:2.34-36}
\begin{split}
x_0=x_i+U_it=x_j+U_jt-\xi_{ij}(0); \\ y_0=y_i=y_j+\eta_{ij}; \quad z_0=z_i=z_j,
\end{split}
\end{equation}
where 
\begin{equation} \label{eq:2.37}
\xi_{ij}(t)=x_j-x_i=(U_i-U_j)t+\xi(0).
\end{equation}
where $\xi_{ij}(0)$ is the initial longitudinal distance between bodies $i$ and $j$. The flow about the $i$--th body is considered asymptotically steady, and can be estimated by standard methods of slender body theory \citep{Newman:1977}. It is also assumed that the separation distance $\eta_{ij}$ is $O(\epsilon L_i)$ to allow calculations of small lateral separation distances. Thus, the three dimensional velocity potential in outer region is expanded in a \textit{Taylor} series about the other body. Using the method of asymptotic expansions we find a solution to the longitudinal motion. The inner solution is governed by the two--dimensional \textit{Laplace} equation and the no penetration boundary condition. The outer solution is governed by the three dimensional \textit{Laplace} equation and by the condition at infinity where the potential diminishes. These solutions are matched in an overlap region, leading to, after rather long but straight forward algebra, expressions for the longitudinal and lateral forces, and moment acting on body $j$ due to the presence and/or movement of body $i$ \citep{Newman:1977, Kadri:2005, Kadri:2014}:
\small
\begin{equation} \label{eq:2.81}
X_j=\sum\limits_{i=1}^n\frac{\rho }{4\pi}\int\limits_{L_i}S'_i(x_i)\left[U_i^2 +U_j^2\int\limits_{L_j}S'_i(x_i){T_j(x_j)\sigma_{ij}\dd x_j}\right] \dd x_j\dd x_i,
\end{equation}
\begin{equation} \label{eq:2.63}
Y_j=\sum\limits_{i=1}^n\frac{\rho U_j\eta_{ij}}{4\pi} \int\limits_{L_i}  (2U_j-U_i)S'_i(x_i) \int\limits_{L_j}T_j(x_j) \dd x_j \dd x_i,
\end{equation}
\begin{equation} \label{eq:2.67}
\begin{split}
N_j=\sum\limits_{i=1}^n\frac{\rho U_j\eta_{ij}}{4\pi}\int\limits_{L_i} \left[x_i(2U_j-U_i)S'_i(x_i)+2U_jS_i(x_i)\right]  
 \int\limits_{L_j}T_j(x_j)\dd x_j\dd x_i \dd x_i.
\end{split}
\end{equation}
\normalsize
where $n$ is the school size, $j=1,2,...,n$, and
\[T_j(x_j)={S'_j(x_j)}{\left(\sigma_{ij}^2+\eta_{ij}^2\right)^{-3/2}} \]
\[\sigma_{ij}=(x_j-x_i-\xi_{ij}); \quad S_j(x_j)=S_j(0)\left(1-{4x_j^2}/{L_j^2}\right)\]
where $S_j(x_j)$ is chosen to be a simple sectional area distribution of parabolic form; $S_j(0)$ is a constant related to the cross-sectional area of the $j$-th body and $S_j(x_j)={\pi r_j^2}$; and $r_j$ is the radius of the cross-sectional area. For nondimensional representation we define 
\[F_{X_j}\equiv{X_jL^2}/{\rho U^2S^2}; \quad
F_{Y_j}\equiv{Y_jL^2}/{\rho U^2S^2}; \quad M_{N_j}\equiv{N_jL}/{\rho U^2S^2}. \]
\subsection{Data collection and analysis} 
A total of 48 photos and eight videos (total duration: 300 seconds) of fish schools from 11 different species were examined for analysis. Data presented in Fig. 2 was processed from a movie (MPEG-4 format) recorded movement and distribution of fish in a school for 00.59.36 min. The video was converted to TIF file formats (at 15 frames per second giving a total of 894 frames) using the tool iMovie (Mac). Positions (x and y coordinates) of fish were determined at three different frames (71, 86 and 104), corresponding to movement at times 00.04.00, 00.05.44 and 00.06.59 min, respectively, using image analysis software (SigmaScan Pro 5.0). 
\subsection{Numerical calculations}
In the case of Figs. 3 and 4, the velocities and lengths of the fish were considered unity, and the slenderness parameter $\epsilon=0.1$. For the diamond pattern cases the longitudinal and lateral distances between each two neighbouring fish rows and columns are $\xi_0=1.1$, and $\eta_0=0.12$, respectively. In the case of a random pattern mode, the same amount of fish were randomly distributed within a similar domain size; the longitudinal and lateral distances were calculate based on a {\it Monte Carlo} algorithm as presented in the statistical guidelines. In the case of Fig. 5, the algorithm was extended to three-dimensions.
\subsection{Statistical guidelines}
The random pattern data presented in Fig. 3 were obtained by carrying out a {\it Monte Carlo} algorithm. Each data point represents an average of repeated random computations of a size of at least a hundred repetitions. For each school size, $n\times n$, the length and width of the computation domain, $\mathbf{l}\times\mathbf{w}$, are given by $\mathbf{l}=n\times L_i$, and $\mathbf{w}=n\times d$. The location of the fish are generated randomly, such that no overlaps are allowed. The longitudinal and later forces, and yawing moments between each two fish are computed using Eqs. (\ref{eq:2.81}), (\ref{eq:2.63}), and (\ref{eq:2.67}). Note that the random pattern data presented in subplots (a)-(j) of Fig. 4, are for a single calculation (no repetition). The (layer) school size is $20 \times 20$.
\begin{figure} \label{fig:graphs}
\centering{ \epsfig{figure=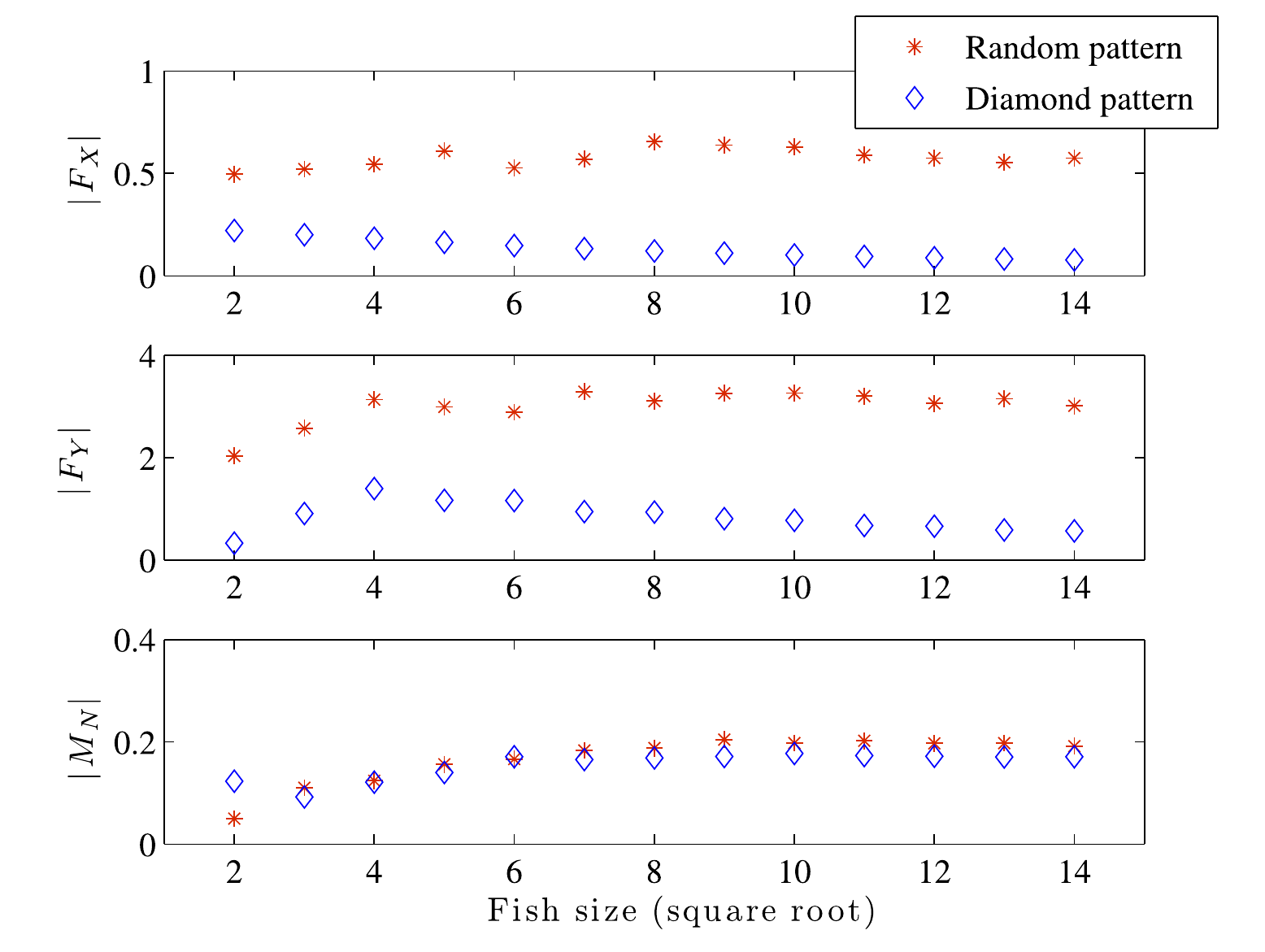,width=\linewidth}}
\caption{Nondimensional hydrodynamic forces and moments as function of fish (square root) school size; for random ($\ast$) and diamond ($\diamond$) patterns. {\bf Top}: longitudinal forces. {\bf Middle}: lateral forces. {\bf Bottom}: yawing moments} 
\end{figure}
\section{Results}
The analysis presented here considers two structure mode patterns, diamond and random. The mean longitudinal and lateral forces acting on a fish in a random pattern mode is larger than those in a diamond mode, whereas the mean moments are similar in both modes (Fig. 3). In this respect, a diamond-shaped swimming pattern is optimal in terms of energy saving which supports previous findings by \cite{Weihs:1983}. Such mode might be observed in schools migrating in `safe' zones, or in large fish or mammals, e.g. dolphins \citep{Weihs:2004, Kadri:2005}, that use the saved energy for extra thrust during escape. However, smaller fish counts on their manoeuvrability for survival, which increases with the total hydrodynamic forces \citep{Wu:1981, Liu:2011}. Table 1 compares between the mean total hydrodynamic forces, $F_{tot}=\sqrt{F_X^2+F_Y^2}$, of the two patterns. It indicates that the mean total force in random patterns is two to five times larger than in diamond patterns. 
\begin{figure} \label{fig:distribution}
\centering{ \epsfig{figure=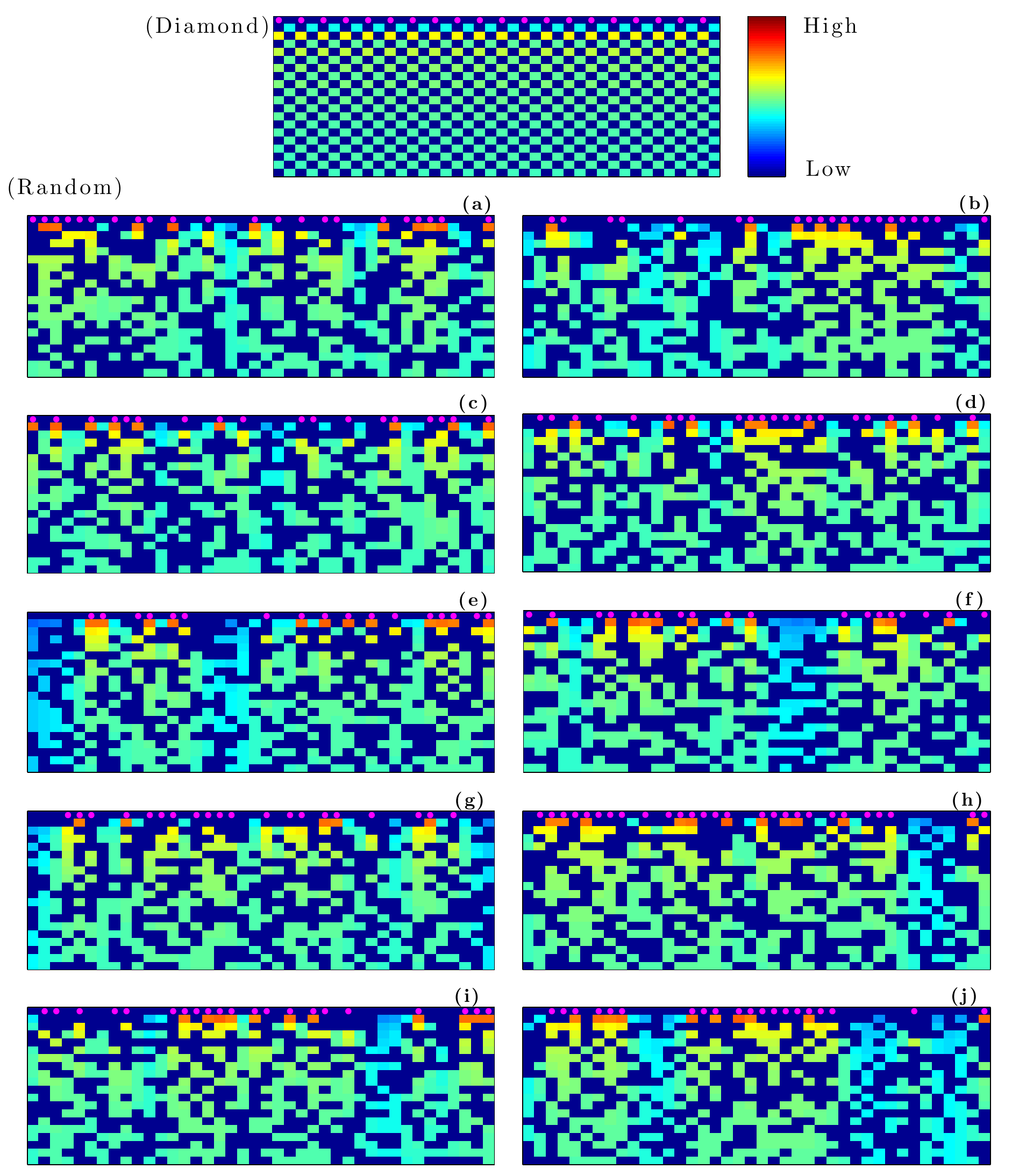,width=\linewidth}}
\caption{The hydrodynamic force effect of the upper fish row (presented by {\color{magenta}$\bullet$}) on the remaining fish school. interaction intensity: blue (low) to red (high). Upper subplot: diamond pattern. Remaining subplots: random patterns.} 
\end{figure}
\begin{table}
  \begin{center}
\def~{\hphantom{0}}
  \begin{tabular}{ccc}
      School size  & Diamond & Random\\[3pt]
\hline
       $2\times2  $ & $0.3974  $&$2.0754$ \\
       $3\times3  $ &  $0.9302 $& $2.6132$\\
       $4\times4  $ &   $1.4072$& $3.1733$\\
       $5\times5  $ &   $1.1768$& $3.0591$\\
       $6\times6  $ &   $1.1690$& $2.9400$\\
       $7\times7  $ &   $0.9532$& $3.3220$\\
       $8\times8 $  &   $0.9422$& $3.1863$\\
       $9\times9$   &   $0.8139$& $3.3187$\\
       $10\times10$   & $  0.7812$& $3.3280$\\
       $11\times11$   &   $0.6810$& $3.2530$\\
       $12\times12 $  &   $0.6641$& $3.1084$\\
       $13\times13$  &   $0.5921$& $3.1943$\\
       $14\times14$   &  $ 0.5734$& $3.0777$\\
  \end{tabular}
  \caption{Calculations of the mean total forces per school size for diamond and random patterns. In the case of a random pattern, the mean total forces are factor of two to five times larger.}
  \label{tab:trenches}
  \end{center}
\end{table}
Swimming in a random pattern mode enables fish to interact more intensely with remote fish members, resulting in a faster and more efficient `information' transfer, from a hydrodynamic perspective. This can be easily seen by the following example. Assume a (layered) rectangular fish school of the size of $20\times 20$ (Fig. 4), the fish on the sides of the rectangle represent an envelope that separates the remaining fish from the surrounding. If , for the sake of brevity, the whole upper fish row (presented by ({\color{magenta}$\bullet$}) encounters a danger, then the survivability of the whole school depends on how fast this information is transferred through the whole school, again from a hydrodynamic perspective. In other words, we are interested in the distribution of the total hydrodynamic effect of the first fish row on the remaining fish school. In a diamond pattern mode (upper subplot) the effect on each row is almost homogeneous. While the lateral forces experienced by the second row are relatively small, due to the fact that $\xi_{ij}$ is large, the effect is largest on the third row and the general trend is that the information (hydrodynamic interaction) decreases with the (double) rows. However, for random pattern modes (subplots (a)-(j)) the information penetrates through the rows, which can be seen in the figure by the differences in colour gradients across the vertical layers (i.e. orange yellow and green compared to blue). Thus, the reaction at the next time instant would occur at multilevel rows simultaneously, which enhances the overall manoeuvrability of the fish as a school.

\begin{figure} \label{fig:3D}
\centering{ \epsfig{figure=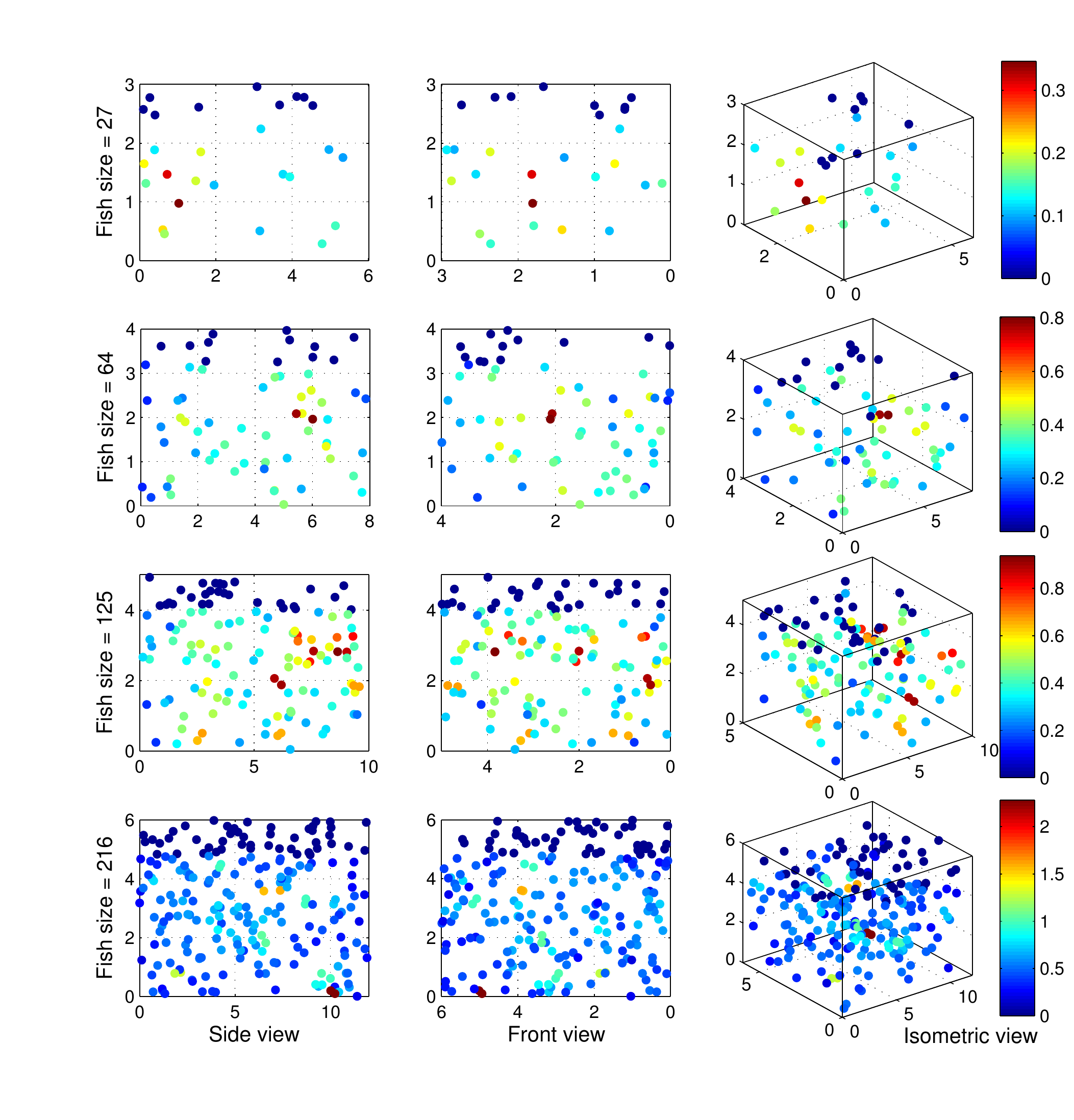,width=\linewidth}}
\caption{The hydrodynamic force effect of the upper fish layer (presented by dark blue {\color{darkblue}$\bullet$}) on the remaining fish school. interaction intensity: blue (low) to red (high). {\bf Left}: side views (fish move to the right).  {\bf Middle}: front views (fish move into page). {\bf Right}: Isometric view. First row: 27 fish, in a $6\times3\times3$ box. Second row: 64 fish, in a $8\times4\times4$ box. Third row: 125 fish, in a $10\times5\times5$ box. Fourth row: 216 fish, in a $12\times6\times6$ box.} 
\end{figure}
\section{Discussion}
Within a random pattern a manoeuvring fish experiences, on average, larger centripetal forces, and thus can reach larger angular velocities ($\omega \propto {F_{tot}}^{1/2}$). Since each fish, within a random pattern, applies on average larger hydrodynamic forces on the school, its manoeuvring, e.g as a response to danger, results in larger impact on the fish school and in particular on its proximate neighbours. Therefore, the hydrodynamic changes within the school as a whole are larger in case of random pattern mode, which enhances its survivability. Note that the mathematical analysis presented in Fig. 4 considers discrete layers of the school, an assumption which is rarely met in nature \citep{Breder:1965, Oshima:1950, Partridge:1979}. However, it is easy to show that the three dimensional fish school analysis would result in larger hydrodynamic forces, which in turn, further enhances the school manoeuvrability, and alertness. Such an analysis is carried out in Fig. 5. Here, we examined the three-dimensional effect of the upper fish group (presented in dark blue {\color{darkblue}$\bullet$}) on the remaining school members. We considered four different school sizes, $27$, $64$, $125$, and $216$, within boxes of dimensions  $6\times3\times3$, $8\times4\times4$, $10\times5\times5$, and $12\times6\times6$, respectively. The fish swim from left to right relative to the side view. It is notable here that the hydrodynamic effects made by the upper fish group on the remaining school members is somewhat disordered, which can be seen by the inhomogeneous distribution of colours. In reality, the analyses of the videos show that the different fishes, of each school, may have different lengths, speeds, and orientations, as well as locomotion techniques, which would all add to the disordered behaviour of transferring the hydrodynamic ``information'' among the school members, both spatially and temporally. 
 
The analysis made here has been described in the context of specific fish schooling species, though similar analysis can be carried out for other fishes, swarming behaviour in general, and bird flocking in particular, e.g. by a straight forward extension of the work by \cite{Higdon:1978}, and \cite{Higdon:1975}. These support the general hypothesis that a disordered and non-trivial collective behaviour of individuals within a nonlinear dynamical system is essential for optimising transfer of information - an optimisation that might be crucial for survival. The work presented here can also be applied for interaction between multiple AUV's with a submarine, e.g. \cite{Leong:2015}.

\subsubsection*{Acknowledgments.} 
The authors gratefully acknowledge Martina and Herbert Bauder for providing the photos and videos.  

\bibliographystyle{elsart-harv}

\end{document}